\documentclass[amsmath,amssymb,prc,final,showpacs,twocolumn]{revtex4-1} 

\usepackage{bm,hyphenat,xspace,draftcopy}
\usepackage{graphicx,epsfig}

\newcommand{\scl}{0.63}

\newcommand{\Ref}{Ref.}

\newcommand{\cm}{\mathrm{c\!\:\!.m\!\:\!.}}

\begin{document}

\title {Momentum-space calculation of proton-deuteron scattering including
Coulomb  and irreducible three-nucleon forces}

\author{A.~Deltuva} 
\email{deltuva@cii.fc.ul.pt}
\affiliation{Centro de F\'{\i}sica Nuclear da Universidade de Lisboa, 
P-1649-003 Lisboa, Portugal }

\received{October 26, 2009}


\begin{abstract}
Three-nucleon scattering equations with irreducible three-nucleon force 
are solved in momentum-space.
The Coulomb interaction between the two protons is included 
using the method of screening and renormalization.
The need for the renormalization of the scattering amplitudes
is demonstrated numerically.
The Coulomb and Urbana IX three-nucleon force effects on the observables
of elastic proton-deuteron scattering and breakup are studied.
\end{abstract}

 \maketitle

\section{Introduction \label{sec:intro}}

Nucleon-deuteron scattering has been extensively studied in the past
with the aim of testing various nuclear interaction models,
i.e., nucleon-nucleon $(NN)$ potentials and  also three-nucleon $(3N)$ 
forces.
While the theoretical description of  neutron-deuteron $(nd)$ scattering
is considerably simpler and a number of highly sophisticated calculations 
\cite{witala:01a,kuros:02b,deltuva:03c,doleschall:04a,ishikawa:07a}
exists with realistic $NN$  potentials and $3N$ forces, the experimental data
are much more abundant and precise for proton-deuteron $(pd)$ reactions.
In that case there are only few realistic calculations above 
the deuteron breakup threshold. Configuration-space treatment based on the
Kohn variational principle \cite{kievsky:01a,kievsky:04a}
uses the local Argonne $V_{18}$ (AV18) $NN$ potential \cite{wiringa:95a}
and the irreducible Urbana IX (UIX) $3N$ force \cite{urbana9} 
but is limited so far
to elastic scattering at energies below 50 MeV  in the center-of-mass (c.m.)
system; only very recent work \cite{ishikawa:09a}
solving integral Faddeev equations in configuration-space provides $pd$ 
breakup results at proton lab energy 13 MeV. 
In contrast, the momentum-space treatment \cite{deltuva:05a,deltuva:05c} based 
on the screening and renormalization method \cite{taylor:74a,alt:78a} 
has lead to $pd$ elastic scattering and breakup
results for energies up to pion-production threshold;
this method can be used with
nonlocal $NN$ potentials as well, but so far only an
effective $3N$ force due to the virtual excitation of a nucleon 
to a $\Delta$ isobar \cite{deltuva:03c} has been included. 
The aim of the present work is to overcome
that limitation, i.e., to extend the technique of 
Refs.~\cite{deltuva:05a,deltuva:05c} to include also an irreducible $3N$ force.

In Sec.~\ref{sec:th} we derive three-particle scattering equations 
including the three-body force.
In Sec.~\ref{sec:coulomb} we discuss the inclusion of the Coulomb
interaction  using the method of  screening and
renormalization and demonstrate its validity numerically.
Section~\ref{sec:res} presents some characteristic effects of Coulomb
and  Urbana IX  $3N$ force in  $pd$ elastic scattering and breakup.
Section~\ref{sec:concl} gives our summary.

\section{Three-particle scattering equations including irreducible 
three-body force  \label{sec:th}}

For the description of three-particle scattering interacting via 
three pairwise potentials $v_\alpha$, $\alpha=1, 2, 3$,
we used Alt, Grassberger, and Sandhas (AGS) equations \cite{alt:67a}. 
In this section we give a short derivation of the AGS equations
including a three-body force
\begin{equation} 
 V_{(3)} = \sum_{\alpha=1}^3 u_\alpha
\end{equation}
which is decomposed into three terms $u_\alpha$ that are symmetric in the
exchange of particles $\beta \neq \alpha$ and  $\gamma \neq \alpha$
and can be transformed into one another by a cyclic permutation;
these properties will be used later for the symmetrization of equations.
 The full resolvent is
\begin{subequations} 
\begin{align} \label{eq:G}
G = {} & (E+i0 - H_0 - \sum_\gamma v_\gamma - \sum_\gamma u_{\gamma})^{-1}, \\
G = {} & G_\beta + G_\beta \sum_\gamma ( \bar{\delta}_{\beta \gamma} v_\gamma 
+ u_\gamma) G, \label{eq:Gg}
\end{align}
\end{subequations}
where $\bar{\delta}_{\beta \alpha} = 1 - {\delta}_{\beta \alpha}$,
$E$ is the available three-particle energy and $H_0$ the three-particle 
kinetic energy operator, both in the center-of-mass (c.m.) system, and 
the channel resolvents are
\begin{subequations} 
\begin{align} \label{eq:Gc}
G_\gamma = {} & (E+i0 - H_0 - v_\gamma )^{-1}, \\
G_\gamma = {} & G_0 +  G_0 T_\gamma G_0.
\end{align}
\end{subequations}
Here $G_0 = (E+i0 - H_0)^{-1}$ is the free resolvent and 
\begin{equation}  \label{eq:T}
 T_\gamma =  v_\gamma + v_\gamma G_0 T_\gamma
\end{equation}
the two-particle transition matrix.

The  multichannel three-particle transition operators $U_{\beta \alpha}$
are defined by the decomposition of the  full resolvent into channel resolvents
according to
\begin{equation} \label{eq:GU}
G =  \delta_{\beta \alpha} G_\alpha + G_\beta U_{\beta \alpha} G_\alpha .
\end{equation}
Inserting Eq.~\eqref{eq:GU} with $\beta=\gamma$ into  Eq.~\eqref{eq:Gg},
comparing it back to Eq.~\eqref{eq:GU} and taking into account 
Eqs.~\eqref{eq:Gc} and \eqref{eq:T} we obtain integral equation
for the multichannel three-particle transition operators
\begin{gather} \label{eq:Uba}
\begin{split}
 U_{\beta \alpha} = {} & \bar{\delta}_{\beta \alpha} G_0^{-1} + u_\alpha
+ \sum_\gamma  \bar{\delta}_{\beta \gamma} T_\gamma G_0 U_{\gamma \alpha} \\
 & + \sum_\gamma  u_\gamma G_0(1+T_\gamma G_0)  U_{\gamma \alpha}.
\end{split}
\end{gather}
The on-shell matrix elements
$\langle\phi_{\beta}|U_{\beta \alpha}|\phi_{\alpha}\rangle$
are  amplitudes (up to a factor) for  elastic 
($\beta = \alpha$), rearrangement ($0 \neq \beta \neq \alpha$) scattering,
and breakup ($\beta=0$). 
The channel states $|\phi_{\alpha}\rangle$  are the
eigenstates of the corresponding channel Hamiltonian $H_\alpha = H_0 + v_\alpha$
with the energy eigenvalue $E$ while  $|\phi_{0}\rangle$
describes the free relative motion of three particles; the dependence on the
Jacobi momenta and discrete quantum numbers is suppressed in our notation.

For nucleon-deuteron scattering it is convenient to 
 consider nucleons as identical particles in the
isospin formalism where the symmetrized amplitudes are
$ \sum_\alpha \langle\phi_{\beta}|U_{\beta \alpha}|\phi_{\alpha}\rangle$.
 The symmetrized transition operator for elastic scattering
is the solution of the symmetrized AGS integral equation
\begin{subequations} \label{eq:AGSs}
\begin{gather} 
\begin{split}
U = {} & PG_0^{-1} + (1+P)u + PTG_0U \\ & + (1+P)uG_0(1+TG_0)U.
\end{split}
\end{gather}
We omit the spectator index that is not needed anymore. The basis states are
antisymmetric in the pair only, the full antisymmetry is ensured
by $P = P_{12}P_{23} + P_{13}P_{23}$ where $P_{\alpha \beta}$ is  the
permutation operator of particles $\alpha$ and  $\beta$. 
The breakup operator is then obtained from the quadrature
\begin{gather} 
U_0 = (1+P)[G_0^{-1} +u +TG_0U + uG_0(1+TG_0)U].
\end{gather}
\end{subequations}
For the practical solution it is convenient to introduce the transition operator
\begin{gather} 
X = G_0^{-1} +u +TG_0U + uG_0(1+TG_0)U
\end{gather}
such that
\begin{subequations} \label{eq:U-X}
\begin{align} 
U = {} & [P + u G_0 (1+P)] X, \\
U_0 = {} & (1+P) X.
\end{align}
\end{subequations}
Operator $X$ is obtained from the integral equation 
\begin{gather} \label{eq:X}
X = G_0^{-1} + TG_0PX + (1+TG_0)uG_0(1+P)X.
\end{gather}
The practical advantage of this equation is that all terms have the same 
structure with respect to the relative momentum of the interacting pair in
the final state; this is convenient for the interpolation that is needed
 to calculate $PX$ at each iteration step. 
The handling of the $3N$ force is discussed in the Appendix~\ref{sec:u9}.
Otherwise, the numerical technique for solving Eq.~\eqref{eq:X} 
in momentum-space partial-wave representation
and calculating on-shell elements in Eqs.~\eqref{eq:U-X} is taken over
from Refs.~\cite{deltuva:03a,deltuva:phd}. There, two equally reliable
interpolation methods for the two-nucleon transition matrices were used.
However, in the presence of the irreducible $3N$ force we use the one based on 
splines that is more convenient than the one based on the Chebyshev expansion.

Finally we note that in Refs.~\cite{witala:01a,kuros:02b} 
nucleon-deuteron scattering 
equations of similar form were solved. The relation between our 
transition operator $X$ and the operator $T$  of 
 Refs.~\cite{witala:01a,kuros:02b},
 not to be confused with the two-nucleon transition matrix $T$ in the 
equations above, reads $X = G_0^{-1} + T$.

\section{Inclusion of the Coulomb force \label{sec:coulomb}}

In order to include the Coulomb interaction 
we use the method of screening
and renormalization \cite{taylor:74a,alt:78a}
as described in detail in Refs.~\cite{deltuva:05a,deltuva:05d,deltuva:08c}
for pairwise interactions. The screened Coulomb potential $w_{\gamma R}$,
that in the configuration space representation has the form
$w_{\gamma R}(r) = (\alpha_e/r)\exp(-(r/R)^n)$, $\alpha_e \approx 1/137$ being
the fine structure constant,
is added to the hadronic proton-proton $(pp)$ potential $v_\gamma$.
The $pp$ transition matrix \eqref{eq:T} is calculated
with the full interaction $v_{\gamma} + w_{\gamma R}$ and used to solve
the AGS equations yielding multichannel transition operators
$ U^{(R)}_{\beta \alpha}$  ($ U^{(R)}$ and $ U^{(R)}_{0}$ in the symmetrized
version);  their dependence on the Coulomb screening radius $R$ 
is indicated. 
Following the strategy of Refs.~\cite{deltuva:05a,deltuva:05d,deltuva:08c} 
it is straightforward to decompose the amplitudes into long-range 
and Coulomb-distorted short-range parts also when the $3N$ force is present. 
The only difference is that
the integral equations for the reduced short-range operators 
$ \tilde{U}^{(R)}_{\beta \alpha}$ 
of  Refs.~\cite{deltuva:05a,deltuva:05d,deltuva:08c} 
contain additional three-body force terms, i.e.,
\begin{subequations} 
  \begin{align} \nonumber
 \tilde{U}^{(R)}_{\beta\alpha} = {} &
\bar{\delta}_{\beta \alpha} (G_{\alpha R}^{-1} + v_{\alpha}) + u_\alpha + 
{\delta}_{\beta \alpha} \mathcal{W}_{\alpha R}  \\ &
+ \sum_\gamma (\bar{\delta}_{\beta \gamma} v_\gamma + u_\gamma +
{\delta}_{\beta \gamma} \mathcal{W}_{\beta R})
G_{\gamma R} \tilde{U}^{(R)}_{\gamma\alpha}, \\
    \tilde{U}^{(R)}_{0\alpha} = {} &
    G_{\alpha R}^{-1} + v_{\alpha}  + u_\alpha
    + \sum_\gamma (v_\gamma + u_\gamma) G_{\gamma R} \tilde{U}^{(R)}_{\gamma\alpha},
\end{align}
\end{subequations}
where $G_{\alpha R}$ and $\mathcal{W}_{\alpha R}$ are defined in
 Refs.~\cite{deltuva:05a,deltuva:05d,deltuva:08c}.
Nevertheless, the relation of $ \tilde{U}^{(R)}_{\beta \alpha}$
to  the full AGS operators $ U^{(R)}_{\beta \alpha}$ as well as 
the screened Coulomb contributions diverging in the  $R \to \infty$ limit
remain the same. Thus, the renormalization prescription can be taken
over from  Refs.~\cite{deltuva:05a,deltuva:05d,deltuva:08c}
where the amplitudes for $pd$ elastic scattering and breakup referring to
 unscreened Coulomb are calculated as
\begin{subequations} \label{eq:UC}
\begin{align} 
\langle \phi' |U^{(C)} |\phi \rangle  = {} & \nonumber
    \langle \phi' |T^{\cm}_C |\phi \rangle \\ + &    \label{UC1}
    \lim_{R \to \infty}     \{ \mathcal{Z}_{R}^{-1}(q_i)
\langle \phi' | [U^{(R)}- T^{\cm}_R] |\phi \rangle \}, \\ \label{UC0}
\langle \phi_0 |U_0^{(C)} |\phi \rangle  = {} &
\lim_{R \to \infty} \{ z_R^{-\frac12}(p_f) \langle \phi_0 |U_0^{(R)} |\phi \rangle 
\mathcal{Z}_{R}^{-\frac12}(q_i) \}.
\end{align}
\end{subequations}
The long-range part of the screened elastic scattering amplitude 
is given by the two-body on-shell transition matrix 
$\langle \phi' |T^{\cm}_{R}| \phi \rangle$
derived from the screened  Coulomb potential between proton
and the c.m. of the deuteron.
Renormalized by  $\mathcal{Z}_{R}^{-1}(q_i)$
in the $R \to \infty$ limit it converges
(in general, as a distribution) to the well-known pure Coulomb amplitude
$\langle \phi' |T^{\cm}_{C}| \phi \rangle$. 
The Coulomb-distorted short-range parts  
$\langle \phi' | [U^{(R)}- T^{\cm}_R] |\phi \rangle$ and 
$\langle \phi_0 |U_0^{(R)} |\phi \rangle$  are calculated numerically
at finite $R$ since after the renormalization they rapidly
converge with $R$ due to their short-range nature;
 we only have to make sure that $R$ is large enough for the desired accuracy.
In Eqs.~\eqref{eq:UC} $q_i$ is the magnitude of the initial relative $pd$ 
momentum and $p_f$  is the magnitude of the final relative $pp$ momentum.
The renormalization factors $\mathcal{Z}_{R}(q_i)$ and
$z_R(p_f)$ are diverging phase factors 
defined in \Ref~\cite{taylor:74a} for a general screening
and given in Refs.~\cite{deltuva:05a,deltuva:05d,deltuva:08c}
for the form of screening used there as well as in this work.
We get well-converged results for the observables when the 
Coulomb-distorted short-range part of the amplitudes is calculated
using the screening function with $n=4$ and  $R=10$ fm  (30 fm)
for $pd$ elastic scattering (breakup).
With those values of $R$ the  partial-wave expansion 
converges slower than in the $nd$ case but the AGS equations still can be
 solved  in the partial-wave basis. 
In our first calculations with Coulomb \cite{deltuva:05a,deltuva:05d} 
high two-baryon partial waves needed for the convergence were included using 
the perturbative approach of Ref.~\cite{deltuva:03b} that was confirmed to
be highly reliable in later calculations where we found an efficient 
method \cite{deltuva:06a} to include the high partial waves exactly.
This latter method  \cite{deltuva:06a} is used in the present
work where we obtain fully converged results by taking into account
the hadronic interaction in two-nucleon partial waves with
total pair angular momentum $I \le 5$ and
the screened Coulomb interaction in two-proton partial waves
with pair orbital angular momentum $L \le 14$. The partial waves with
total $3N$ angular momentum  $\mathcal{J} \le \frac{59}{2}$ are considered
but it is fully sufficient to include the $3N$ force only in those with
$\mathcal{J} \le \frac{19}{2}$.
Of course, both total $3N$ isospin $\mathcal{T}=\frac12$ and $\frac32$
states are included.

Thus, the presence of the $3N$ force does not change the rate of the 
$R$-convergence for the observables whose detailed study was given in 
Refs.~\cite{deltuva:05a,deltuva:05d} and will not be repeated here.
However, a recent work \cite{witala:09a}
on an alternative Coulomb treatment in $pd$ scattering 
proposed a  renormalization prescription
that is different from ours given in Eqs.~\eqref{eq:UC}.
 According to Ref.~\cite{witala:09a}, the
 $pd$ elastic scattering amplitude calculated with screened Coulomb
does not need renormalization at all, i.e., the limit
$\lim_{R \to \infty} \langle \phi' |U^{(R)}|\phi \rangle$ should exist. 
The renormalization for the breakup amplitude is needed but it is different 
from ours given in Eq.~\eqref{UC0}. 
However, numerical results of Ref.~\cite{witala:09a}
involve approximations that are not well under control: the screened
Coulomb transition matrix is approximated by the screened
Coulomb potential and, furthermore, particular contributions to the 
 $pd$ scattering amplitudes that include  first order terms  
in the screened Coulomb transition matrix are neglected.
Therefore we feel the need to clarify the issue of renormalization.
We study the dependence on the screening radius $R$
for selected components of the
nonrenormalized and renormalized $pd$ elastic scattering
amplitudes, $\langle \phi' | U^{(R)} | \phi \rangle$ and 
$\langle \phi' | U^{(C)} | \phi \rangle$,
calculated at proton lab energy $E_p = 9$ MeV; 
the AV18 $NN$ potential \cite{wiringa:95a} and the Urbana IX $3N$ 
force \cite{urbana9} are taken as the hadronic interaction.
In Fig.~\ref{fig:URZs} we show spin-nondiagonal amplitudes that have
only the Coulomb-distorted short-range part.
The nonrenormalized amplitude 
$\langle \phi' | U^{(R)} | \phi \rangle$ shows a clear $R$-dependence;
in fact, its absolute value becomes $R$-independent but the phase is 
proportional to $\ln R$. In contrast,
the renormalized amplitude $\langle \phi' | U^{(C)} | \phi \rangle$,
within the accuracy of the plot, becomes independent of $R$ for $R \ge 10$ fm.
The renormalized spin-diagonal amplitude
that has also a long-range part becomes $R$-independent as well, 
as Fig.~\ref{fig:UZl} demonstrates. Thus, fully converged numerical results
without uncontrolled approximations
clearly support the standard screening and renormalization theory as given in 
Eqs.~\eqref{eq:UC} and not the one of Ref.~\cite{witala:09a}.

\renewcommand{\scl}{0.64}
\begin{figure*}[!]
\begin{center}
\includegraphics[scale=\scl]{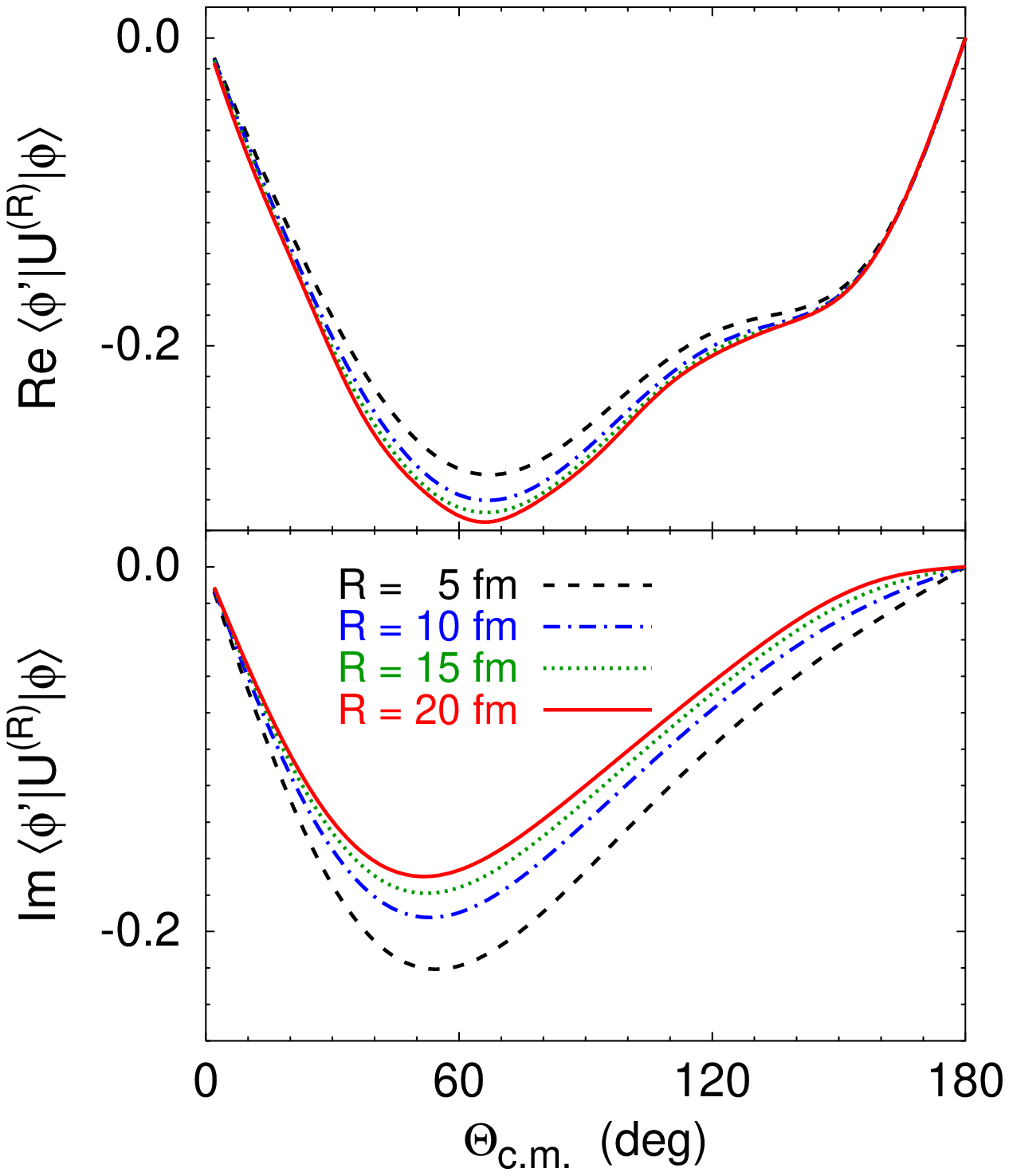} \hspace{2mm}
\includegraphics[scale=\scl]{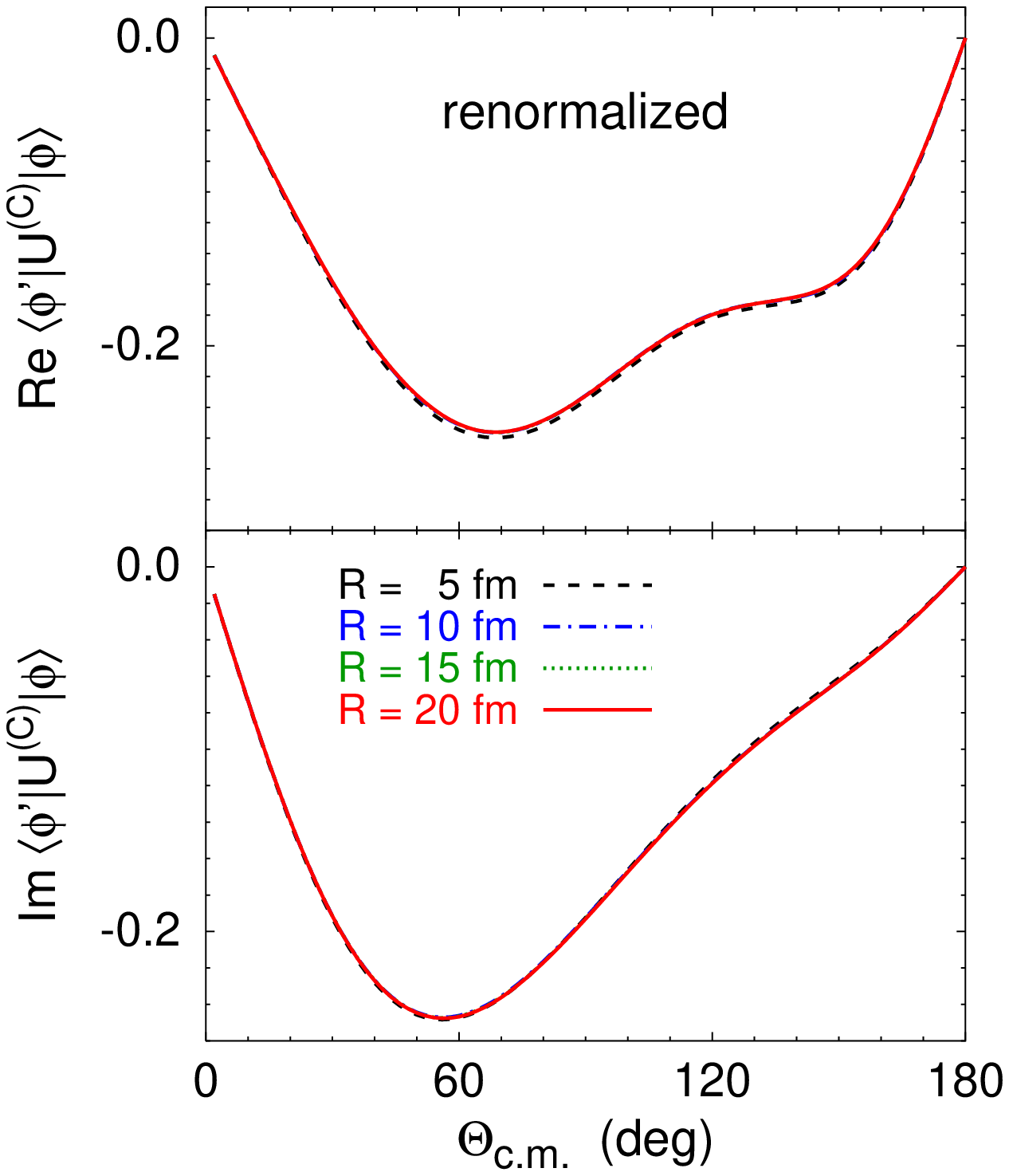}
\end{center}
\caption{\label{fig:URZs} (Color online)
Real and imaginary parts of 
nonrenormalized (left side) and renormalized (right side) $pd$ elastic 
scattering amplitudes (in arbitrary units) at $E_p = 9$ MeV 
shown as functions of the c.m. scattering angle.
The initial and final particle spin projection
quantum numbers are $m_p=\frac12$, $m'_p=-\frac12$, and  $m_d = m'_d = 1$.
Results for the AV18 + UIX force model
 obtained with the screening radius 
$R= 5$~fm (dashed curves), 10~fm (dash-dotted curves), 
15~fm (dotted curves), and 20~fm (solid curves) are compared.}
\end{figure*}

\renewcommand{\scl}{0.64}
\begin{figure}[!]
\begin{center}
\includegraphics[scale=\scl]{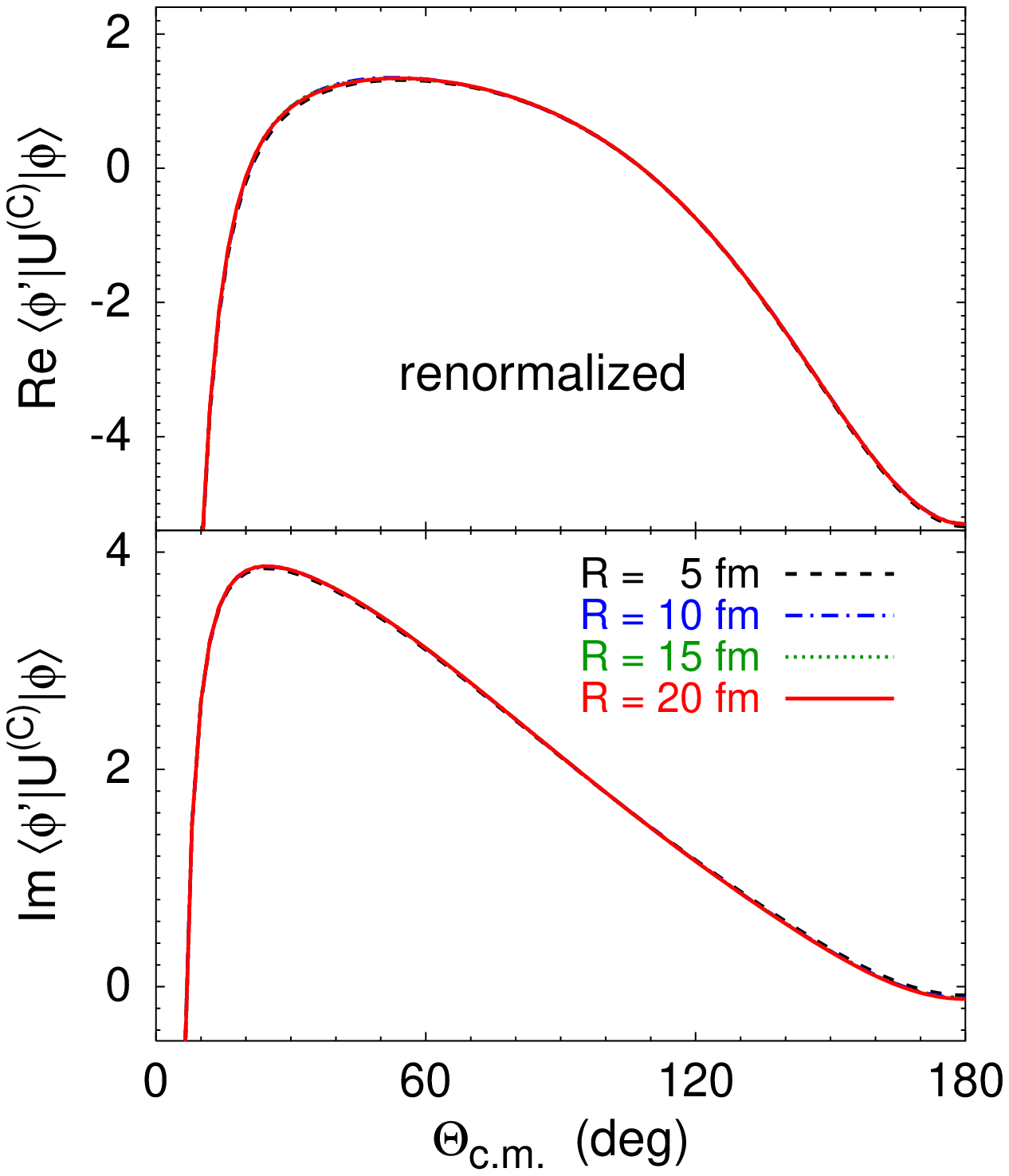}
\end{center}
\caption{\label{fig:UZl} (Color online)
Real and imaginary parts of the renormalized  $pd$ elastic 
scattering amplitude. 
The initial and final particle spin projection
quantum numbers are $m_p=m'_p=\frac12$,  and  $m_d = m'_d = 1$.
Curves as in Fig.~\ref{fig:URZs}.}
\end{figure}

\section{Results \label{sec:res}}

Numerical results of this paper are derived from the
AV18 $NN$ potential \cite{wiringa:95a}
together with the Urbana IX $3N$ force \cite{urbana9}; 
this combination, AV18 + UIX,
is one of the most widely used $NN$ + $3N$ force models.
Hadronic charge dependence is fully included, i.e.,
$pp$ and $np$ potentials are used within the isospin formalism.
Point Coulomb interaction is added for $pp$ but
the additional electromagnetic terms of AV18 are not included.
In order to isolate Coulomb and $3N$ force effects, calculations
without Coulomb or without the $3N$ force are performed as well.
Obviously, we have many more predictions than it is possible and wise
to show. Therefore we make a judicious selection and present 
the  most interesting cases.
The readers, dissatisfied with our choice, are welcome to obtain the results
for their favorite data from us.

\begin{figure*}[!]
\begin{center}
\includegraphics[scale=0.58]{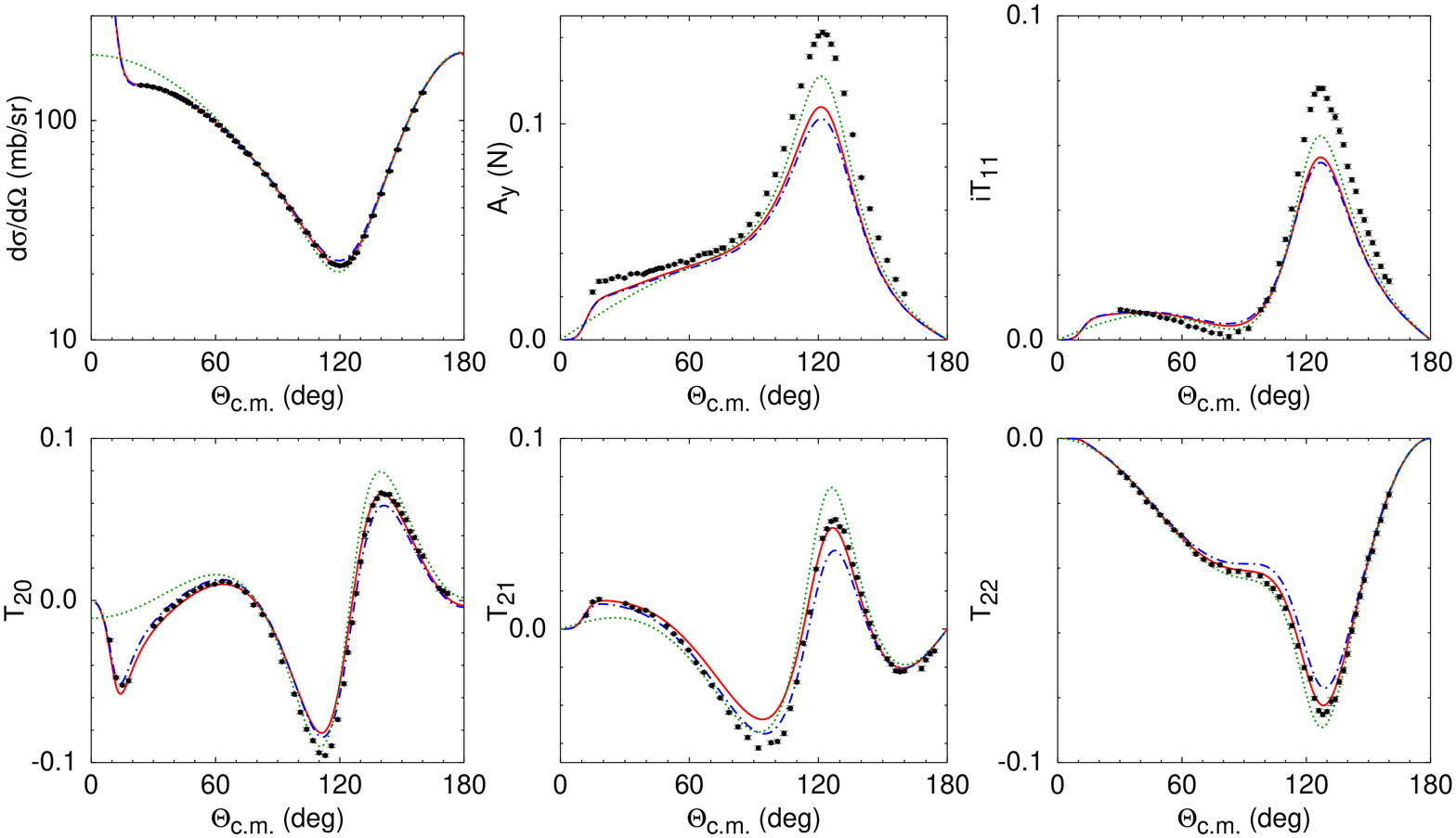}
\end{center}
\caption{\label{fig:pd9} (Color online)
Differential cross section and analyzing powers for $pd$ elastic scattering
at  9~MeV proton lab energy as functions of the c.m. scattering angle.
Results for AV18 (dashed-dotted curves) and AV18 + UIX (solid curves)
force models, both including Coulomb, and for AV18 + UIX without Coulomb
(dotted curves) are compared with the 
 experimental data from Ref.~\cite{sagara:94a}.}
\end{figure*}

In Fig.~\ref{fig:pd9} we show results for the differential cross section 
and analyzing powers of $pd$ elastic scattering
at  9~MeV proton lab energy (18~MeV deuteron lab energy). 
The Coulomb effect is important in the whole kinematical regime. 
There are well-known discrepancies \cite{kievsky:01a}
around the maximum of the vector analyzing powers $A_y(N)$ and $iT_{11}$
and the first minimum of $T_{21}$.
The $3N$ force effect is rather small; it is mostly a scaling effect
due to the change in the binding energies of ${}^3\mathrm{H}$ and
 ${}^3\mathrm{He}$ that are calculated in the Appendix~\ref{sec:EB}.
 In addition, observables shown in Fig.~\ref{fig:pd9} serve as a benchmark
since they already have been calculated in Ref.~\cite{kievsky:01a} 
using AV18 and AV18 + UIX force models together with the Coulomb force. 
The $3N$ Schr\"odinger equation was solved in configuration space
using the Kohn variational
principle and explicitly imposing the proper Coulomb boundary conditions.
The $pd$ elastic scattering results obtained with the
two methods for including the $pp$ Coulomb interaction,
i.e., our momentum-space screening and renormalization method
and the one of  Ref.~\cite{kievsky:01a},
were compared in Ref.~\cite{deltuva:05b} taking the AV18 $NN$ potential
as the hadronic interaction alone; 
a good agreement over a wide range of energies
for all studied observables was found. Comparing  Fig.~\ref{fig:pd9} to the
9~MeV results of  Ref.~\cite{kievsky:01a} it is easy to see that the
predicted $3N$ force effect is the same in both cases. 
Thus, the agreement between the momentum-space and configuration-space
results including both the  $3N$ force and
Coulomb and  is as good as without the $3N$ force.

\begin{figure*}[!]
\begin{center}
\includegraphics[scale=0.58]{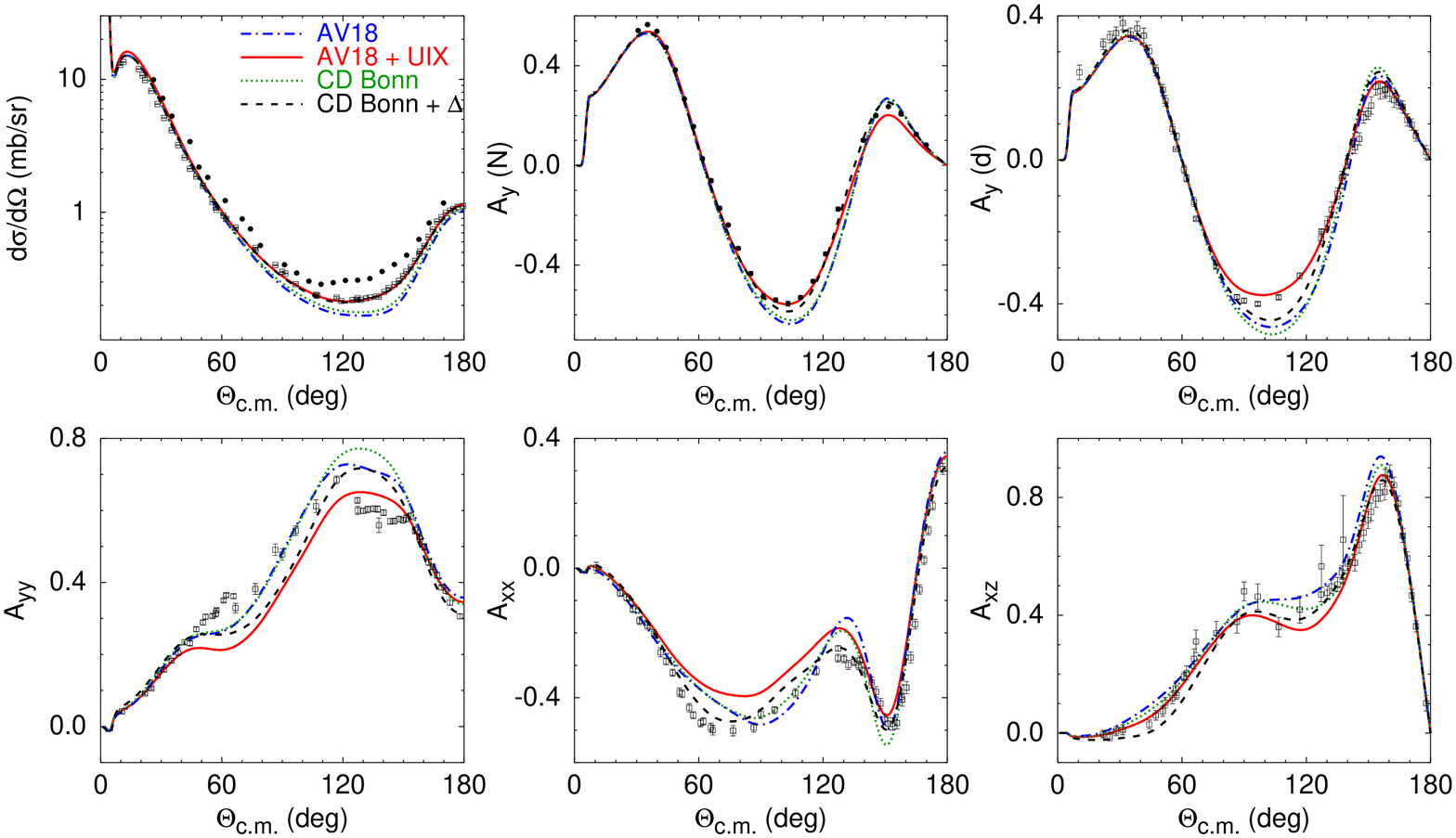}
\end{center}
\caption{\label{fig:pd135} (Color online)
Differential cross section and analyzing powers for $pd$ elastic scattering
at  135~MeV proton lab energy.
Results for AV18 (dashed-dotted curves), AV18 + UIX (solid curves),
CD Bonn (dotted curves), and CD Bonn + $\Delta$ (dashed curves) force models,
all including Coulomb, are compared with the 
 experimental data from Ref.~\cite{ermisch:05a} (full circles)
and Ref.~\cite{sekiguchi:02a} (open squares).}
\end{figure*}

In Fig.~\ref{fig:pd135} we show  the differential cross section 
and analyzing powers for $pd$ elastic scattering
at  135~MeV proton lab energy (270~MeV deuteron lab energy)
where  the configuration-space calculations are not available so far.
As found in Ref.~\cite{deltuva:05a} and confirmed in the present work, 
the Coulomb effect at this relatively high energy is large only at 
forward angles.
We therefore do not show it separately. Instead, since the $3N$ force
effect is quite significant, it is interesting to compare 
two different $3N$ force models: (a) Urbana IX, based on 
the Fujita-Miyazawa force \cite{fujita:57a}, i.e., 
the two-pion $(2\pi)$ exchange with an intermediate $\Delta$-isobar excitation,
and supplemented by a purely phenomenological repulsive short-range part;
(b) an effective $3N$ force due to explicit $\Delta$-isobar excitation
that uses no static approximation for the propagation of the $\Delta$,
 includes beside the pion also the exchange of heavier mesons $\rho$, $\omega$,
and $\sigma$, and has higher order contributions, e.g.,
three-meson ring diagrams; all contributions are consistent 
with each other since they are built from the same two-baryon
coupled-channel potential CD Bonn + $\Delta$ \cite{deltuva:03c}
that is as realistic as its purely nucleonic reference potential
CD Bonn \cite{machleidt:01a}. The $\Delta$-isobar effect is isolated
as the difference between predictions of the CD Bonn + $\Delta$ and CD Bonn 
potentials. In most cases both  Urbana IX and $\Delta$-isobar effects are
qualitatively similar although the former is usually larger, especially
around the minimum  of the deuteron vector analyzing power $A_y(d)$
where the predictions of AV18 + UIX are considerably closer to the 
experimental data. The reason why the $\Delta$-isobar effect is smaller,
at least partially, is  the $NN$ dispersion that is
generated by the explicit $\Delta$-isobar excitation
and often competes with the effective $3N$ force  \cite{deltuva:03c}.
The two $3N$ force models show qualitatively different behavior
around the first minimum of the deuteron tensor analyzing power
$A_{xx}$ where the CD Bonn + $\Delta$ accounts for the data better.
Finally we note that our AV18 + UIX results without Coulomb (not shown here) 
are consistent with the corresponding results of Ref.~\cite{witala:01a}.

\begin{figure*}[!]
\begin{center}
\includegraphics[scale=0.62]{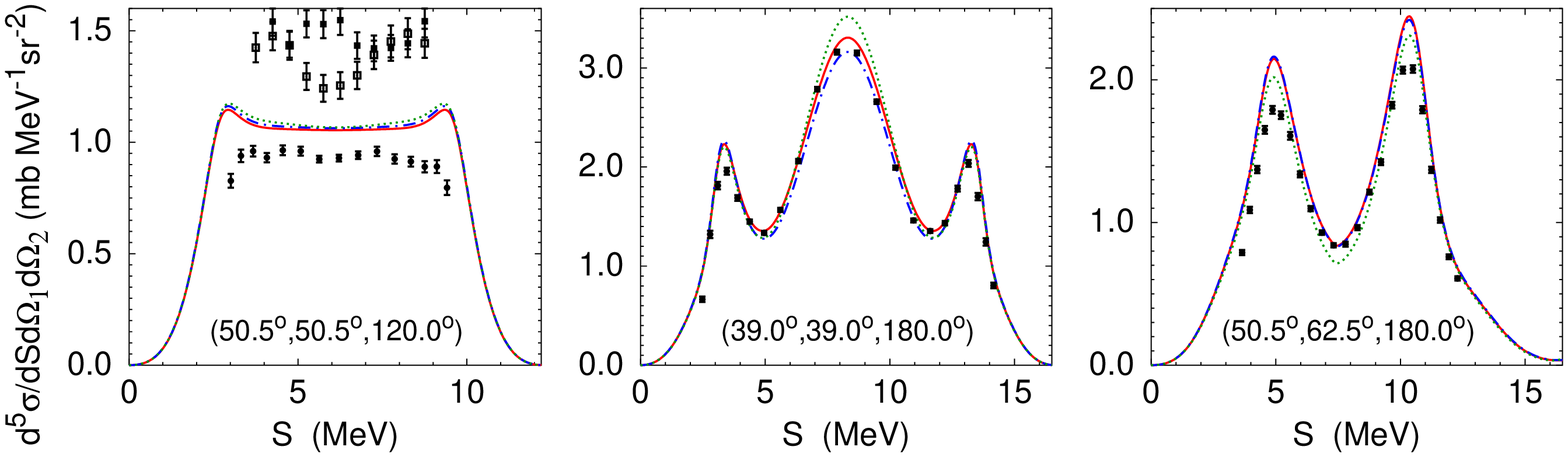}
\end{center}
\caption{\label{fig:pd13} (Color online)
Differential cross section for $pd$ breakup at 13~MeV proton lab energy 
in space star (left), quasifree scattering (middle), and
collinear (right) configurations 
as function of the arclength $S$ along the kinematical curve.
Curves as in Fig.~\ref{fig:pd9} and  the 
 experimental data from Ref.~\cite{rauprich:91} (full circles).
For the space star configuration  also the $nd$ data from 
Refs.~\cite{strate:89,setze:05a} (open and full squares)
are shown.}
\end{figure*}

In Fig.~\ref{fig:pd13} we show the fivefold differential cross section
for $pd$ breakup at 13~MeV proton lab energy in few special
kinematical configurations that are characterized in a
standard way by the final-state
polar angles of the two detected protons and by the
azimuthal angle between them, 
$(\theta_1, \theta_2, \varphi_{12} = \varphi_2 - \varphi_1)$.
Though the inclusion of Coulomb slightly improves the agreement with
data in the space star configuration, the Coulomb effect is far too small
to reproduce the difference between the experimental $pd$ and $nd$ data and
to resolve the so-called \emph{space star anomaly}.
Slightly larger and beneficial Coulomb effects are seen in quasifree 
scattering (QFS) and collinear configurations; the  differential cross section
is decreased around QFS peak and increased around the collinear point (minimum) 
and $np$ final-state interaction (FSI) peaks. The remaining discrepancies
around the $np$-FSI peaks, at least to some extent, 
may be due to the finite geometry, not taken into account in our 
calculations owing to the lack of information on experimental details.
The $3N$ force effect is very small except for the QFS configuration.
All these findings are consistent with our previous results
\cite{deltuva:05d} derived from the CD Bonn + $\Delta$ and CD Bonn 
potentials.

\begin{figure*}[!]
\begin{center}
\includegraphics[scale=0.62]{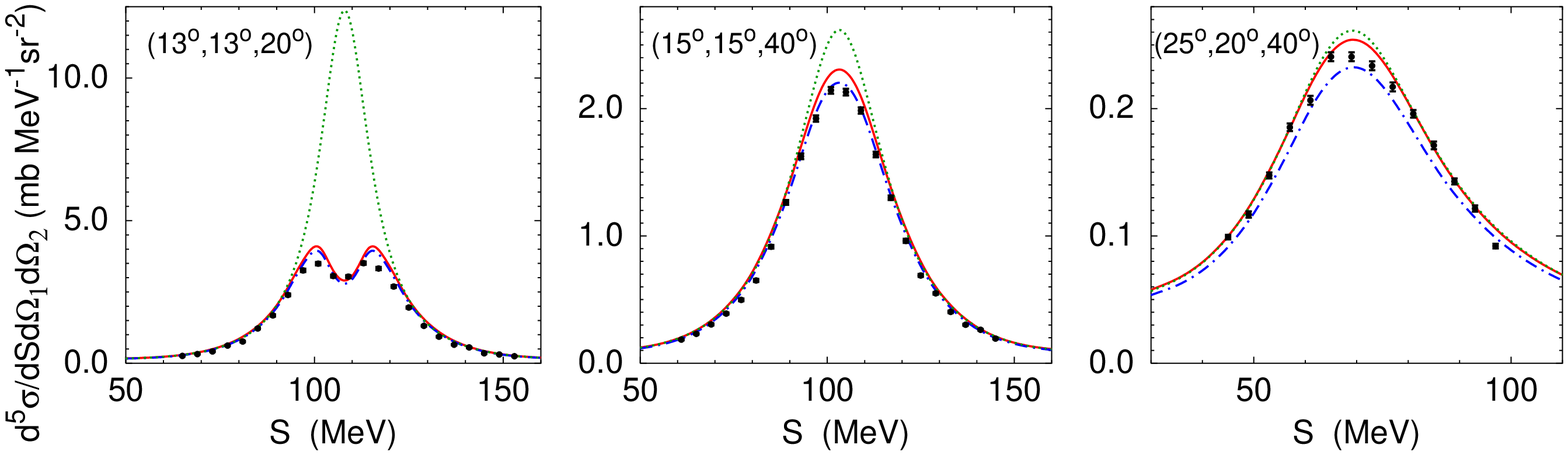}
\end{center}
\caption{\label{fig:dp130_40} (Color online)
Differential cross section for $pd$ breakup
at  130~MeV deuteron lab energy in selected kinematical configurations
with small relative azimuthal angle.
Curves as in Fig.~\ref{fig:pd9} and  the 
 experimental data from Refs.~\cite{kistryn:05a,kistryn:06a}.}
\end{figure*}

\begin{figure*}[!]
\begin{center}
\includegraphics[scale=0.62]{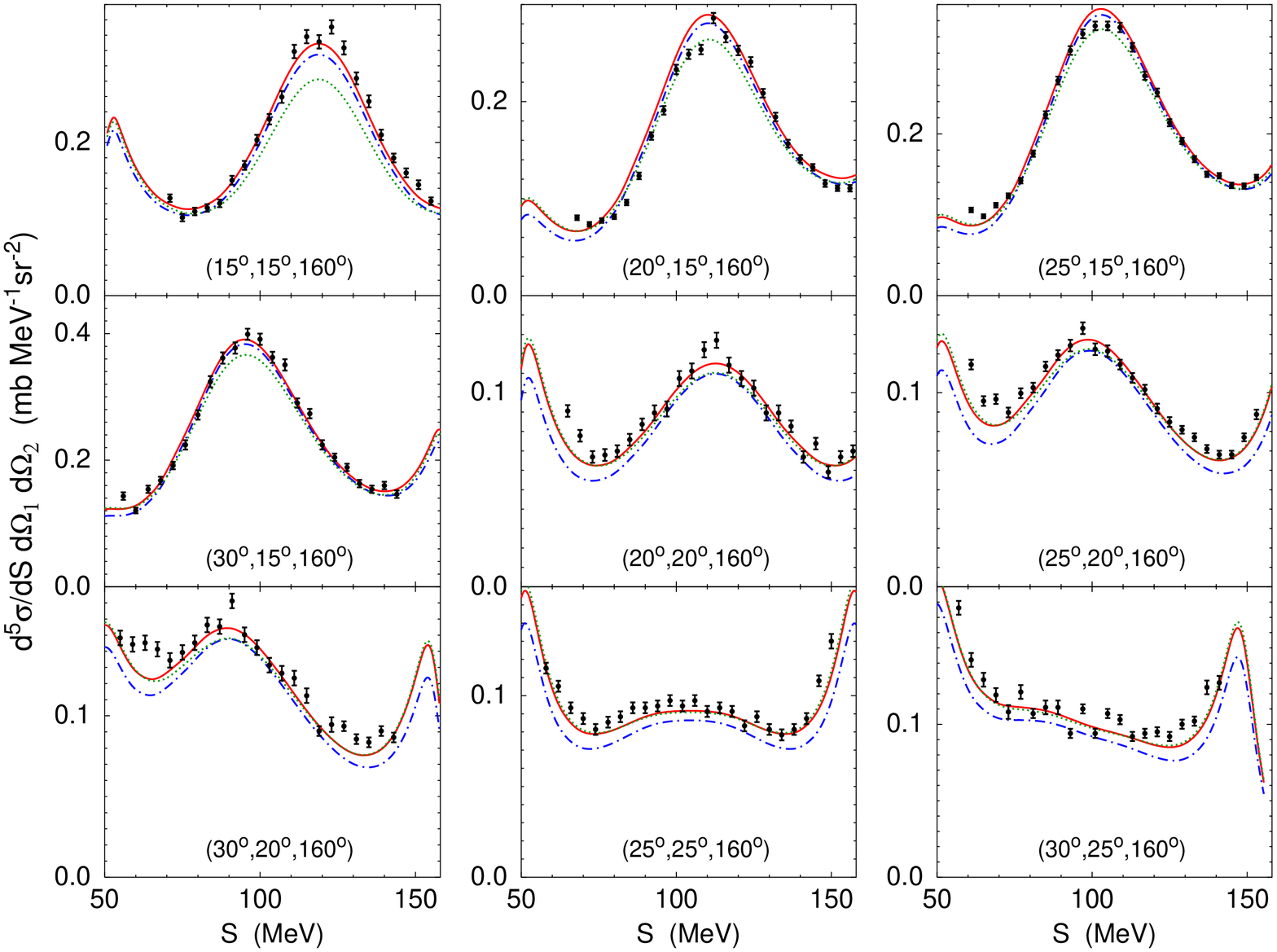}
\end{center}
\caption{\label{fig:dp130_160} (Color online)
Differential cross section  for $pd$ breakup
at  130~MeV deuteron lab energy in selected kinematical configurations
with large relative azimuthal angle.
Curves as in Fig.~\ref{fig:pd9} and  the 
 experimental data from Ref.~\cite{kistryn:05a}.}
\end{figure*}

\begin{figure*}[!]
\begin{center}
\includegraphics[scale=0.62]{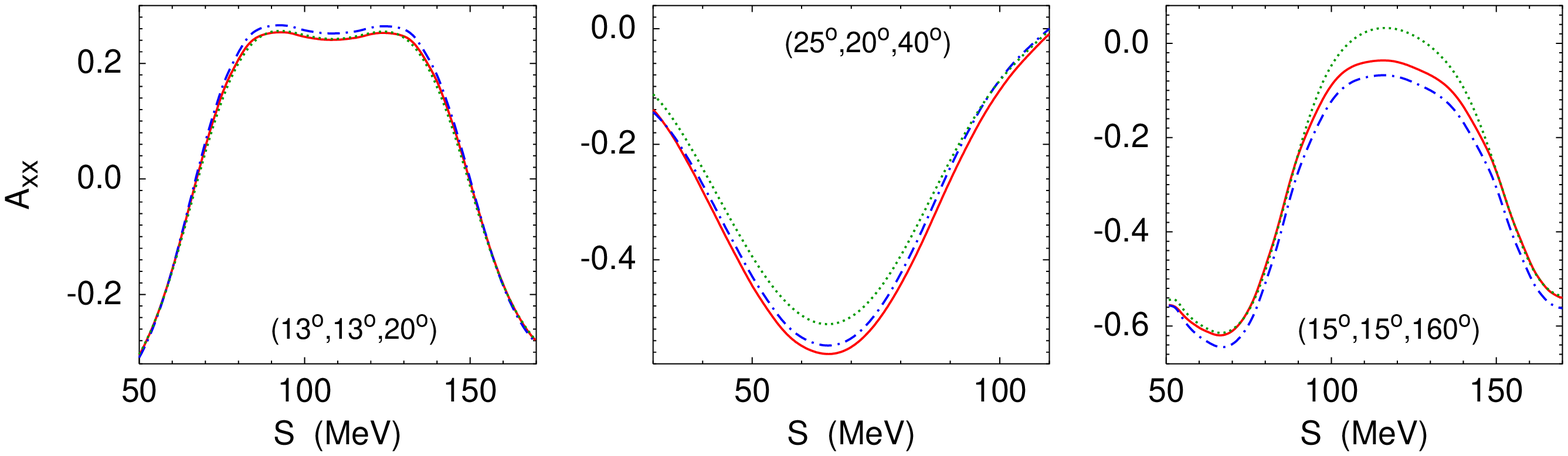}
\end{center}
\caption{\label{fig:dp130_axx} (Color online)
Deuteron analyzing power $A_{xx}$ for $pd$ breakup
at  130~MeV deuteron lab energy in selected kinematical configurations.
Curves as in Fig.~\ref{fig:pd9}.}
\end{figure*}

Finally, we consider $pd$ breakup  at 130 MeV deuteron lab energy 
that was measured recently in a  variety of kinematical configurations 
\cite{kistryn:05a,kistryn:06a}. 
In some of them we found sizable Coulomb effects for the differential
cross section \cite{deltuva:05d,kistryn:06a} but very small 
$\Delta$-isobar effects. In contrast, the present calculations 
as well as those without Coulomb given in Ref.~\cite{kistryn:05a} reveal
visible effects of the Urbana IX $3N$ force. Therefore
a more extensive study of the interplay between the Coulomb and the
$3N$ force is needed. In Fig.~\ref{fig:dp130_40} we show 
the fivefold differential cross section for few kinematical 
configurations with small relative azimuthal angle  $\varphi_{12}$.
The central point of the $(13^\circ,13^\circ,20^\circ)$ configuration
corresponds to very low relative $pp$ energy in the final state,
$E_{pp} < 0.2$ MeV,  where
the differential cross section and also the $3N$ force effect
are strongly reduced by the Coulomb as a
result of the $pp$ repulsion; the found  Coulomb effect is well supported
by the experimental data.
The  relative $pp$ energy in the final state
increases with the relative polar and azimuthal angles of the protons
and therefore the Coulomb effect decreases
in the remaining configurations of Fig.~\ref{fig:dp130_40}
but the  $3N$ force effect becomes more visible.
The relative $pp$ energy gets larger at  $\varphi_{12} = 160^{\circ}$ 
in  Fig.~\ref{fig:dp130_160}
leading to an increase of the differential cross section due to the Coulomb,
especially in the configurations with smaller $\theta_i$.
Since the total breakup cross section at this energy
 is almost unchanged by Coulomb as demonstrated in Ref.~\cite{deltuva:05d}
and confirmed in the present work, one may expect in particular 
configurations an increase of the differential cross section due to Coulomb
to compensate for the strong decrease in the regions
with low $E_{pp}$. The inclusion of the Urbana IX $3N$ force,
as in  Fig.~\ref{fig:dp130_40}, increases the differential cross section,
especially in the configurations where both $\theta_i$ are large.
Thus, at smaller  $\theta_i$  Coulomb is dominating while
at larger $\theta_i$ the  Urbana IX $3N$ force becomes more significant.
In some kinematical regimes, e.g., around the central peaks
of the configurations $20^\circ \le \theta_1 \le 30^\circ$,
$\theta_2 = 20^\circ$, $\varphi_{12} = 160^{\circ}$, both 
are equally important leading  to quite a satisfactory description
of the data in all studied configurations.

Deuteron analyzing powers were measured in the same experiments
\cite{kistryn:05a,kistryn:06a}, however, the data analysis is not yet completed.
 We therefore present only few examples in  Fig.~\ref{fig:dp130_axx}
demonstrating that for the spin observables like  $A_{xx}$ 
Coulomb and  $3N$ force effects may take place in completely different regions
of the phase space compared to the differential cross section.
E.g., $A_{xx}$ remains unaffected by the Coulomb in the 
$(13^\circ,13^\circ,20^\circ)$ configuration with very low $E_{pp}$ where
a strong decrease of the cross section was found in Fig.~\ref{fig:dp130_40},
but shows a moderate Coulomb effect in the  $(25^\circ,20^\circ,40^\circ)$ 
configuration where it was almost negligible in the case of cross section.
In contrast, the  $3N$ force effect in the latter configuration is only
significant for the differential cross section but not for  $A_{xx}$.
The last  configuration in  Fig.~\ref{fig:dp130_axx} shows moderate
effects of both Coulomb and $3N$ force competing with each other
unlike in the case of
the differential cross section  in Fig.~\ref{fig:dp130_160}.
A detailed study of deuteron analyzing powers in more
kinematical configurations is postponed till the finalization of the 
experimental data that is expected soon  \cite{stephan:09a}.

\section{Summary \label{sec:concl}}

In this paper we derive  AGS integral equations for proton-deuteron
scattering including an irreducible three-nucleon force and solve them in the
momentum-space partial wave representation. 
The Coulomb interaction between the protons is included
using the screening and renormalization method whose validity is 
confirmed by the numerical results for scattering amplitudes;
they are well converged with respect to the
screening and with respect to the quantum number cutoffs.
AV18 $NN$ potential with the Urbana IX $3N$ force are used as the
hadronic interaction model to calculate the
observables of  $pd$ elastic scattering and breakup.
In the low-energy $pd$ elastic scattering, where configuration-space 
calculations \cite{kievsky:01a}
with the same dynamic input are available, a good agreement 
between our results and those of Ref.~\cite{kievsky:01a} is found.
For  $pd$ elastic scattering at higher energies and for $pd$ breakup
we provide first results for the AV18 + UIX force model with Coulomb.
 In higher-energy $pd$ elastic scattering where
the Coulomb effect is confined to small scattering angles but the
$3N$ force effect is significant, we compared two  models, 
Urbana IX  and the effective $3N$ force due to 
explicit $\Delta$-isobar excitation,
 and with few exceptions found a qualitative agreement between them.
In breakup the Coulomb effects are fully consistent with those found
in our earlier calculations \cite{deltuva:05a,deltuva:05d} with different
hadronic interactions. The inclusion of Coulomb is unable to resolve
the space star anomaly at low-energies,
but clearly improves the description of the experimental data
 at 130 MeV deuteron lab energy where, depending on the
kinematical configuration, it may significantly decrease or increase
the differential cross section. A moderate Urbana IX $3N$ force effect
is seen as well that increases the cross section which is only
slightly underpredicted by the theory. A complicated
interplay of Coulomb and $3N$ force effects takes place in the 
spin observables; they will be studied extensively in the future
using other  $3N$ force models as well.

\begin{acknowledgments}
The author thanks R.~Lazauskas for valuable discussions on the inclusion
of the $3N$ force, and A.~C.~Fonseca for the comments on the manuscript.
\end{acknowledgments}

\begin{appendix}

\section{Urbana IX $3N$ force \label{sec:u9}}

The Urbana IX $3N$ force has $2\pi$-exchange and phenomenological 
repulsive short-range terms, 
\begin{gather} \label{eq:U9}
\begin{split}
V_{(3)} =  {} & \sum_{\alpha\beta\gamma \; \mathrm{cyclic}}
\big( A_{2\pi} \{X^\pi_{\alpha\beta},X^\pi_{\beta\gamma}\}
\{\tau_\alpha \cdot \tau_\beta,\tau_\beta  \cdot \tau_\gamma \} \\
& +  C_{2\pi} [X^\pi_{\alpha\beta},X^\pi_{\beta\gamma}]
[\tau_\alpha \cdot \tau_\beta ,\tau_\beta  \cdot \tau_\gamma ] 
 + U_0 T^2_{\alpha\beta} T^2_{\beta\gamma} \big),
\end{split}
\end{gather}
where curly and square brackets denote anticommutator and commutator,
respectively. The strength constants are
$A_{2\pi} = -0.0293$ MeV,  $C_{2\pi} = \frac14 A_{2\pi}$, and
$U_0 = 0.0048$ MeV; the latter should be not confused with the 
symmetrized breakup operator $U_0$ defined in Sec.~\ref{sec:th}.
$\tau_\alpha$ is the isospin vector operator of the nucleon
$\alpha$, and $X^\pi_{\alpha\beta}$ and $T_{\alpha\beta}$
are local potential-like two-nucleon operators depending on the relative 
coordinate of the nucleons $\alpha$ and $\beta$ and, 
in the case of $X^\pi_{\alpha\beta}$, also
on their spins; the explicit expressions can be found in Ref.~\cite{urbana9}.
The transformation of $X^\pi_{\alpha\beta}$ and $T^2_{\alpha\beta}$ to the momentum
space using spherical Bessel functions is straightforward.
For decomposing \eqref{eq:U9} into three symmetric parts we follow
Ref.~\cite{lazauskas:phd}, i.e.,
\begin{gather} \label{eq:u9}
\begin{split}
u_\gamma =  {} & 
 A_{2\pi} X^\pi_{\alpha\beta} \big\{ X^\pi_{\beta\gamma}
[ 2 \tau_\alpha \cdot \tau_\gamma
- \frac{i}{2} \tau_\alpha \cdot (\tau_\beta \times \tau_\gamma) ] \\
& + X^\pi_{\gamma\alpha}
[2\tau_\gamma \cdot \tau_\beta 
+ \frac{i}{2} \tau_\alpha \cdot (\tau_\beta \times \tau_\gamma)] \big\}
\\ &  + \frac12 U_0 T^2_{\alpha\beta} (T^2_{\beta\gamma} + T^2_{\gamma\alpha}),
\end{split}
\end{gather}
$\alpha\beta\gamma$ being cyclic. In the $3N$  scattering  and bound state 
equations, \eqref{eq:X} and \eqref{eq:BS}, 
the $3N$ force component \eqref{eq:u9} always acts on a state $|y \rangle$
that is fully antisymmetric due to $(1+P)$.
Furthermore, taking into account that 
$X^\pi_{\alpha\beta}$,  $T_{\alpha\beta}$, and $u_\gamma$ are symmetric
under exchange of the nucleons $\alpha$ and $\beta$, one can prove that 
certain terms in Eq.~\eqref{eq:u9} yield equal contributions
\cite{lazauskas:phd} and
calculate the desired matrix elements of the $3N$ force component  as
\begin{gather} \label{eq:u9y}
\begin{split}
{}_\gamma\langle \nu | u_\gamma| y \rangle = {} & 
2 A_{2\pi} \sum_{\nu_1 \nu_2 \nu_3}
 {}_\gamma\langle \nu |X^\pi_{\alpha\beta}| \nu_1 \rangle_\gamma \\
& \times  {}_\gamma\langle \nu_1 | [ 2 \tau_\alpha \cdot \tau_\gamma
- \frac{i}{2} \tau_\alpha \cdot (\tau_\beta \times \tau_\gamma) ]
|\nu_2 \rangle_\alpha  \\
& \times  {}_\alpha\langle \nu_2|X^\pi_{\beta\gamma} | \nu_3 \rangle_\alpha
 {}_\alpha\langle \nu_3| y\rangle \\
& +  U_0  \sum_{\nu_1 \nu_2 \nu_3}
 {}_\gamma\langle \nu |T^2_{\alpha\beta}| \nu_1 \rangle_\gamma
 {}_\gamma\langle \nu_1 |\nu_2 \rangle_\alpha  \\
& \times {}_\alpha\langle \nu_2|T^2_{\beta\gamma} |\nu_3 \rangle_\alpha
 {}_\alpha\langle \nu_3| y\rangle ,
\end{split}
\end{gather}
where each of the intermediate states $|\nu_i \rangle_\alpha$,
with $\nu_i$ abbreviating all continuous and discrete quantum numbers
and $\alpha$ denoting the spectator,
is antisymmetric under the exchange of the nucleons $\beta$ and $\gamma$.
The transformation from the basis with the spectator $\alpha$ to the
basis with the spectator $\gamma$ is needed in order to  evaluate
all the matrix elements of $X^\pi_{\alpha\beta}$ and 
$T^2_{\alpha\beta}$  in their proper bases.
Since $|y\rangle$ is fully antisymmetric, 
${}_\alpha\langle \nu_3| y\rangle = {}_\gamma\langle \nu_3| y\rangle $.
Furthermore, ${}_\gamma\langle \nu_1 |\nu_2 \rangle_\alpha =
\frac12 {}_\gamma\langle \nu_1 |P|\nu_2 \rangle_\gamma$,
and ${}_\gamma\langle \nu_1 | [ 2 \tau_\alpha \cdot \tau_\gamma
- \frac{i}{2} \tau_\alpha \cdot (\tau_\beta \times \tau_\gamma) ]
|\nu_2 \rangle_\alpha$ differs from ${}_\gamma\langle \nu_1 |\nu_2 \rangle_\alpha$
only by the isospin factor that can be found in Ref.~\cite{lazauskas:phd}.
Thus, the calculation of the $2\pi$-exchange and short-range terms
needs only one basis transformation for each of them,
and all intermediate states are antisymmetric with respect to the pair.
This constitutes the advantage of the above method \cite{lazauskas:phd}
for handling the Urbana IX $3N$ force
over the one used in Refs.~\cite{witala:01a} that needed two
basis transformations for each term involving, in addition, 
intermediate nonphysical symmetric states.
Finally we note that for handling other types of $3N$ forces
to be used in the future calculations 
the technique proposed in Ref.~\cite{lazauskas:09b} seems to be very
promising.

\section{Three-nucleon bound state \label{sec:EB}}

Starting from the Schr\"odinger equation in the integral form,
\begin{gather}
|\Psi \rangle = G_0 \sum_\alpha (v_\alpha + u_\alpha) |\Psi \rangle,
\end{gather}
decomposing the $3N$ bound-state wave function 
$ |\Psi \rangle = \sum_\alpha |\psi_\alpha  \rangle$ into its Faddeev components
$ |\psi_\alpha\rangle$ with $\alpha = 1,2,3$, and using the definition
of the two-particle transition matrix \eqref{eq:T} it is straightforward
to derive the Faddeev equation 
\begin{gather}
|\psi_\alpha  \rangle = G_0 T_\alpha 
\sum_\beta \bar{\delta}_{\beta \alpha} |\psi_\beta  \rangle
+ (1+G_0 T_\alpha) G_0  u_\alpha \sum_\beta |\psi_\beta  \rangle.
\end{gather}
The symmetrized version of it reads
\begin{gather} \label{eq:BS}
|\psi  \rangle = G_0 T P |\psi \rangle
+ (1+G_0 T) G_0  u (1+P) |\psi \rangle
\end{gather}
where the bound-state  wave function is obtained as 
$|\Psi \rangle =(1+P)|\psi \rangle$.
In practical calculations we solve the Faddeev equation for
$| x \rangle = G_0^{-1}|\psi \rangle $ that has an advantage of having
exactly the same form of kernel as the scattering equation \eqref{eq:X}.
In contrast to scattering calculations, we include hadronic interaction
in two-nucleon partial waves with total pair angular momentum $I \le 6$ and
take into account all electromagnetic terms of the AV18 $NN$ potential.
Our results for binding energies and wave function probabilities
of ${}^3\mathrm{H}$ and  ${}^3\mathrm{He}$ nuclei 
with and without Urbana IX $3N$ force are collected
in Table~\ref{tab:EB}; they are in good agreement with the results
of other groups \cite{nogga:03a,lazauskas:04a}.

\begin{table}[htbp]
  \centering
\begin{ruledtabular}
  \begin{tabular}{l*{6}{c}}
&  $|E_B|$  &  $\langle H_0 \rangle $  &  $P_{S'}$  &
$P_P$  &  $P_D$  &  $P_{\mathcal{T}=3/2}$    \\ \hline
AV18     & 7.621 & 46.72 & 1.292 & 0.066 & 8.509 & 0.0025 \\
AV18 \cite{nogga:03a}(r) & 7.624 & 46.73 & 1.293 & 0.066 & 8.510 & 0.0025 \\
AV18 \cite{nogga:03a}(p) & 7.621 & 46.73 & 1.291 & 0.066 & 8.510 & 0.0025 \\
AV18+UIX & 8.478 & 51.28 & 1.055 & 0.135 & 9.302 & 0.0025  \\
AV18+UIX \cite{nogga:03a}(r) 
         & 8.479 & 51.28 & 1.054 & 0.135 & 9.301 & 0.0025  \\
AV18+UIX \cite{nogga:03a}(p) 
         & 8.476 & 51.28 & 1.052 & 0.135 & 9.302 & 0.0025  \\
\hline
AV18 & 6.923 & 45.68 & 1.526 & 0.065 & 8.466 & 0.0081 \\
AV18 \cite{nogga:03a}(r) & 6.925 & 45.69 & 1.530 & 0.065 & 8.467 & 0.0081 \\
AV18 \cite{nogga:03a}(p) & 6.923 & 45.68 & 1.524 & 0.065 & 8.466 & 0.0081 \\
AV18+UIX & 7.748 & 50.21 & 1.239 & 0.132 & 9.249 & 0.0075  \\
AV18+UIX \cite{nogga:03a}(r) 
         & 7.750 & 50.21 & 1.242 & 0.132 & 9.248 & 0.0075  \\
AV18+UIX \cite{nogga:03a}(p) 
         & 7.746 & 50.21 & 1.235 & 0.132 & 9.248 & 0.0075  \\
 \end{tabular}
\end{ruledtabular}
  \caption{ \label{tab:EB}
Absolute value of binding energy, expectation value of kinetic energy
(both in MeV),
and probabilities of the wave function components (all in \%)
for ${}^3\mathrm{H}$ (top) and  ${}^3\mathrm{He}$ (bottom) nuclei
calculated with  AV18 and AV18+UIX force models.
Results of Ref.~\cite{nogga:03a} obtained using coordinate-space (r)
and momentum-space (p) frameworks are listed as well.}
\end{table}

\end{appendix}

\end{document}